\documentclass{aa520}
\usepackage{graphicx,natbib,epsfig}
\bibpunct{(}{)}{,}{a}{}{,}
\begin{document}
\title{Particle-In-Cell simulations of circularly polarised Alfv\'en wave phase mixing: a new mechanism for electron acceleration
in collisionless plasmas}
\author{David Tsiklauri \inst{1} \and  
Jun-Ichi Sakai \inst{2}\and  Shinji Saito \inst{2}}
\offprints{David Tsiklauri}
\institute{Institute for Materials Research,
School of Computing, Science and Engineering,
University of Salford, Salford, Greater Manchester, M5 4WT, United Kingdom.
 \and  Laboratory for Plasma Astrophysics, Faculty of Engineering, Toyama University,
3190, Gofuku, Toyama, 930-8555,  Japan.}
\date{Received 25 November 2004 / Accepted 21 January 2005}

\abstract{In this work we used Particle-In-Cell simulations to 
study the interaction of circularly polarised Alfv\'en waves with 
one dimensional plasma density inhomogeneities transverse to the uniform magnetic field (phase mixing) 
in collisionless plasmas.
In our preliminary work we reported discovery of a new electron acceleration 
mechanism, in which
progressive distortion of the Alfv\'en wave front, due to the differences in 
local Alfv\'en speed, generates an oblique (nearly parallel to the magnetic
field) electrostatic field. The latter accelerates electrons through the Landau resonance.
Here we report a detailed study of this novel mechanism, including: 
(i) analysis of broadening of the ion
distribution function due to the presence of Alfv\'en waves and (ii) the generation of compressive perturbations due to
both weak non-linearity and plasma density inhomogeneity.
The amplitude decay law in the inhomogeneous regions, 
in the kinetic regime, is demonstrated to be the same as in the MHD approximation 
described by Heyvaerts and Priest (1983).
 }
\titlerunning{PIC simulations of Alfv\'en wave phase mixing...}
\authorrunning{Tsiklauri et al.}
\maketitle

\section{Introduction}
The study of the interaction of Alfv\'en waves (AWs) 
with plasma inhomogeneities is 
important for
both astrophysical and laboratory plasmas.
This is because both AWs and inhomogeneities 
often coexist in these physical systems.
AWs are believed to be
good candidates for plasma heating, energy and momentum transport.
On the one hand, in many physical situations AWs are easily excitable 
(e.g. through convective motion of the solar interior)
and thus they are present in a number of astrophysical systems.
On the other hand, these waves dissipate due to 
the shear viscosity as opposed to
compressive fast and slow magnetosonic waves which dissipate due to the
bulk viscosity.
In astrophysical plasmas shear viscosity is extremely small
as compared to bulk viscosity. Hence,
AWs are notoriously difficult to dissipate.
One of the possibilities to improve AW dissipation is to introduce progressively
decreasing spatial scales, $\delta l \to 0$, into the system (recall that the 
classical dissipation is $\propto \delta l^{-2}$). 
Heyvaerts and Priest have proposed (in the astrophysical context) one such 
mechanism, called 
AW phase mixing \citep{hp83}. It occurs when a linearly polarised
AW propagates in the plasma with a 
one dimensional density inhomogeneity transverse 
to the uniform magnetic field.
In such a situation the initially plane AW front is progressively 
distorted because of different Alfv\'en speeds across the field. 
This creates progressively stronger gradients
across the field (in the inhomogeneous regions 
the transverse scale collapses to zero),
and thus in the case of finite resistivity, dissipation is greatly enhanced.
Hence, it is believed that phase mixing can provide significant plasma
heating. Phase mixing could be also important for laboratory plasmas.
\citet{hc74} proposed the heating of collisionless plasma by 
utilising spatial phase mixing by shear Alfv\'en wave resonance and discussed potential 
applications to toroidal plasma.
A significant amount of work has been done in the context of heating open
magnetic structures in the solar 
corona \citep{hp83,nph86,p91,nrm97,dmha00,bank00,tan01,hbw02,tn02,tna02,tnr03}.
All phase mixing studies so far have been performed in the MHD approximation,
however, since the transverse scales in the AW collapse progressively to zero,
the MHD approximation is inevitably violated. 
This happens when the transverse scale approaches
the ion gyro-radius $r_i$ and then electron gyro-radius $r_e$.
Thus, we proposed to study the phase mixing effect in the kinetic regime, i.e.
we go beyond a MHD approximation. Preliminary results were 
reported in \citet{tss05}, where
we discovered a new mechanism for the acceleration of electrons
due to wave-particle interactions. This has important implications
for various space and laboratory plasmas, e.g. the 
coronal heating problem and acceleration of the solar wind.
In this paper we present a full analysis of the discovered effect
including an analysis of the broadening of the ion
distribution function due to the presence of Alfv\'en waves and the generation of 
compressive perturbations due to both
weak non-linearity and plasma density inhomogeneity.

\section{The model}
We used 2D3V, the fully relativistic, electromagnetic, particle-in-cell (PIC)
code with MPI parallelisation, modified from the 3D3V TRISTAN code \citep{b93}.
The system size is $L_x=5000 \Delta$ and $L_y=200 \Delta$, where
$\Delta(=1.0)$ is the grid size. The periodic boundary conditions for
$x$- and $y$-directions are imposed on particles and fields. There are about
478 million electrons and ions in the simulation. The average number of
particles per cell is 100 in low density regions (see below). 
The thermal velocity of electrons is $v_{th,e}=0.1c$
and for ions is $v_{th,i}=0.025c$.
The ion to electron
mass ratio is $m_i/m_e=16$. The time step is $\omega_{pe} \Delta t=0.05$. Here
$\omega_{pe}$ is the electron plasma frequency.
The Debye length is $v_{th,e}/\omega_{pe}=1.0$. The electron skin depth 
is $c/\omega_{pe}=10 \Delta$, while the ion skin depth is $c/\omega_{pi}=40 \Delta$.
Here $\omega_{pi}$ is the ion plasma frequency.
The electron Larmor radius is $v_{th,e}/\omega_{ce}=1.0 \Delta$, while
the same for ions is $v_{th,i}/\omega_{ci}=4.0 \Delta$.
The external uniform magnetic field, $B_0(=1.25)$,
is in the $x$-direction and the initial
electric field is zero. 
The ratio of electron cyclotron frequency to the electron plasma
frequency is $\omega_{ce}/\omega_{pe}=1.0$, while the same for ions is
$\omega_{ci}/\omega_{pi}=0.25$. The latter ratio is essentially $V_A/c$ -- the Alfv\'en
speed normalised to the speed of light. Plasma $\beta=2(\omega_{pe}/\omega_{ce})^2(v_{th,e}/c)^2=0.02$.
Here all plasma parameters are quoted far away from the density 
inhomogeneity region. The dimensionless (normalised to some reference constant value of $n_0=100$ particles per cell) 
ion and electron density inhomogeneity is described by
\begin{equation}
 {n_i(y)}=
{n_e(y)}=1+3 \exp\left[-\left(\frac{y-100\Delta}{50 \Delta}\right)^6\right]
\equiv F(y).
\end{equation}
This means that in the central region (across the 
$y$-direction), the density is
smoothly enhanced by a factor of 4, and there are the 
strongest density gradients having 
a width of about ${50 \Delta}$ around the 
points $y=51.5 \Delta$ and $y=148.5 \Delta$.
The background temperature of ions and electrons, 
and their thermal velocities
are varied accordingly
\begin{equation}
{T_i(y)}/{T_0}=
{T_e(y)}/{T_0}=F(y)^{-1},
\end{equation}
\begin{equation}
 {v_{th,i}}/{v_{i0}}=
{v_{th,e}}/{v_{e0}}=F(y)^{-1/2},
\end{equation}
such that the thermal pressure remains constant. Since the background magnetic field
along the $x$-coordinate  is also constant, the total pressure remains constant too.
Then we impose a current of the following form
\begin{equation}
{\partial_t E_y}=-J_0\sin(\omega_d t)\left(1-\exp\left[-(t/t_0)^2\right]\right),
\end{equation}
\begin{equation}
{\partial_t E_z}=-J_0\cos(\omega_d t)\left(1-\exp\left[-(t/t_0)^2\right]\right).
\end{equation}
Here $\omega_d$ is the driving frequency which was fixed at $\omega_d=0.3\omega_{ci}$.
This ensures that no significant ion-cyclotron damping is present. Also,
$\partial_t$ denotes the time derivative.
$t_0$ is the onset time of the driver, which was fixed at $50 /\omega_{pe}$
i.e.  $3.125 / \omega_{ci}$. This means that the driver onset time is about 3 ion-cyclotron
periods. Imposing such a current on the system results in the generation of
left circularly polarised AW, which is driven at the left 
boundary of simulation box and has spatial width of $1 \Delta$.
The initial amplitude of the current is such that 
the relative AW amplitude is about 5 \% of the background
(in the low density homogeneous regions),
thus the simulation is weakly non-linear.

\section{Main results}

Because no initial (perpendicular to the external magnetic field) velocity excitation
was imposed in addition to the above specified currents 
(cf. \citet{tn02,dvl01,tt03,tt04}), 
the circularly polarised AW excited (driven) at the left boundary
is split into two circularly polarised AWs which travel in opposite directions. The dynamics of these
waves as well as other physical quantities is shown in Fig.~1.
(cf. Fig.~1 from \citet{tss05} where 
$B_z$ and $E_y$, the circularly polarised Alfv\'en wave
components, were shown for three different times).
A typical simulation, untill $t=875 / \omega_{ce}=54.69 / \omega_{ci}$ takes about 8 days
on the parallel 32 dual 2.4 GHz Xeon  processors.
It can be seen from the figure 
that because of the periodic boundary conditions, a circularly polarised
AW that was travelling to the left, reappeared 
on the right side of the simulation box.
The dynamics of the AW ($B_z,E_y$) progresses in a similar manner as in the 
MHD, i.e. it phase mixes \citep{hp83}.
In other words, the middle region (in $y$-coordinate), i.e. $50 \Delta \leq y \leq 150 \Delta$, travels 
slower because of the density enhancement (note that 
$V_A(y) \propto 1/\sqrt{n_i(y)}$).
This obviously causes a 
distortion of initially plain wave front and the creation of strong gradients
in the regions around $y \approx 50$ and $150$.
In the MHD approximation when resistivity, $\eta$, is finite, 
the AW is strongly dissipated in these regions. This effectively means that the outer and inner parts of the
travelling AW are detached from each other and propagate independently.
This is why the effect is called phase mixing -- after a long time (in the case
of developed phase mixing), 
phases in the wave front become effectively uncorrelated.
Before \citet{tss05}, it was not clear what to expect from our PIC simulation. 
The code is collisionless and there
are no sources of dissipation in it (apart from the 
possibility of wave-particle interactions).
It is evident from Fig.~1 that in the developed stage of phase mixing 
($t=54.69 / \omega_{ci}$), the AW front is substantially damped in the strongest density
gradient regions.
Contrary to the AW ($B_z$ and $E_y$) dynamics we do not see any phase mixing for 
$B_y$ and $E_z$. The latter two behave similarly.
It should be noted that $E_z$ contains both driven (see Eq.(4) above) and 
non-linearly generated fast magnetosonic wave components, with the former being dominant over the
latter. 
Since $B_x$ is not driven, and initially its perturbations are absent,
we see only non-linearly generated slow magnetosonic perturbations confined to the
regions of strongest density gradients (around $y\approx50$ and $150$).
Note that these also have rapidly decaying amplitude.
Also, we gather from Fig.~1 that the density perturbation ($\approx 10$\%)
which is also generated through the weak non-linearity is present too.
These are propagating density oscillations with variation both in overall magnitude (perpendicular to the
figure plane) and across the $y$-coordinate, and they are mainly confined to the strongest density gradients regions
(around $y\approx 50$ and $150$).
Note that dynamics of $B_x,B_y$ and $B_z$ with 
appropriate geometrical switching (because in our
geometry the uniform magnetic field lies along the $x$-coordinate)
is in qualitative agreement with \citet{bank00} (cf their Fig.~9).
The dynamics of remaining $E_x$ component is treated separately in the next figure.
It is the inhomogeneity of the medium $n_i(y)$, $V_A(y)$, i.e.
$\partial / \partial y \not = 0$, is the cause of weakly non-linear coupling of the AWs to the 
compressive modes (see \citet{nrm97,tan01} for further details).

\begin{figure*}
\centering
 \epsfig{file=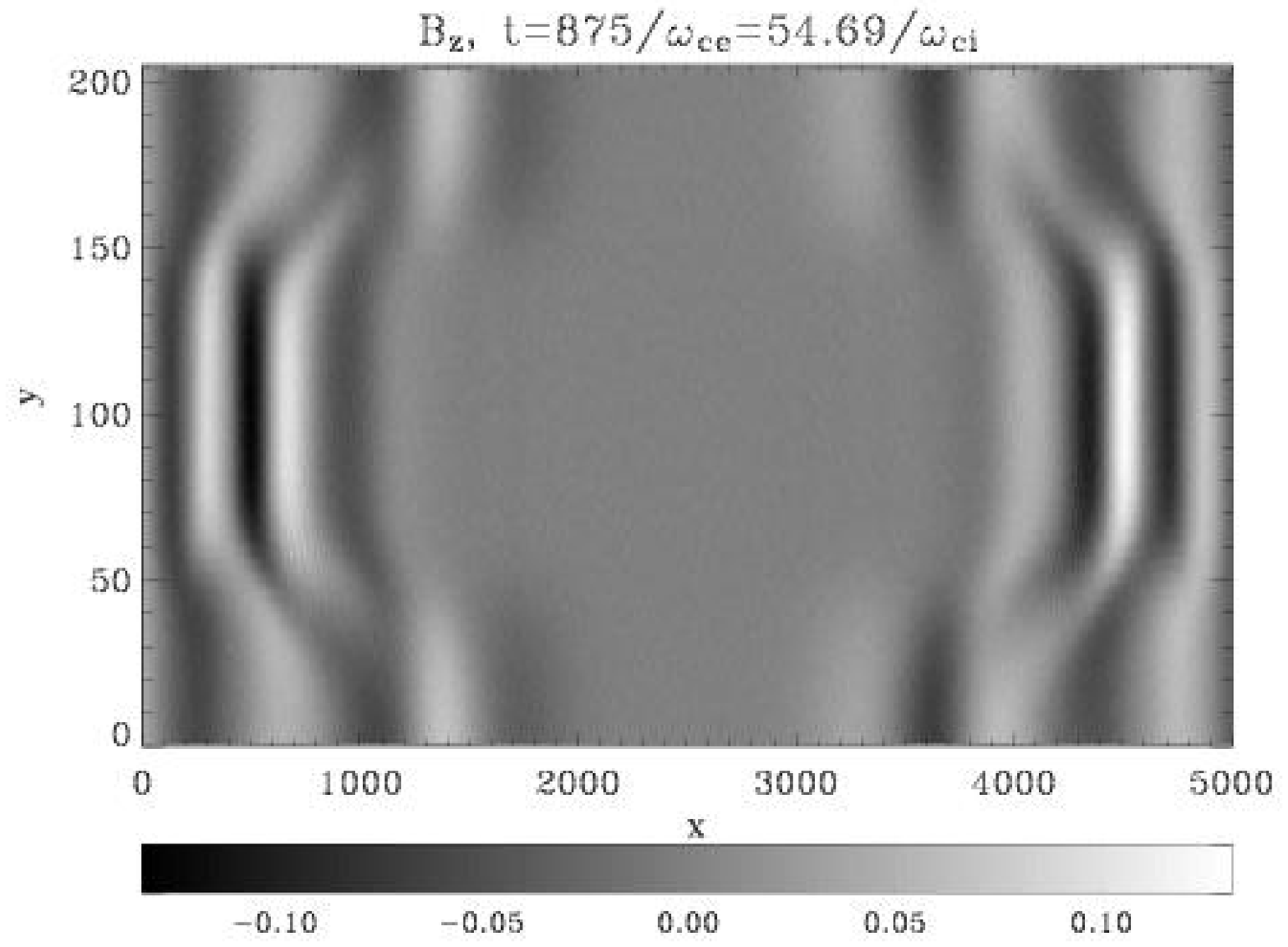,width=6cm}
 \epsfig{file=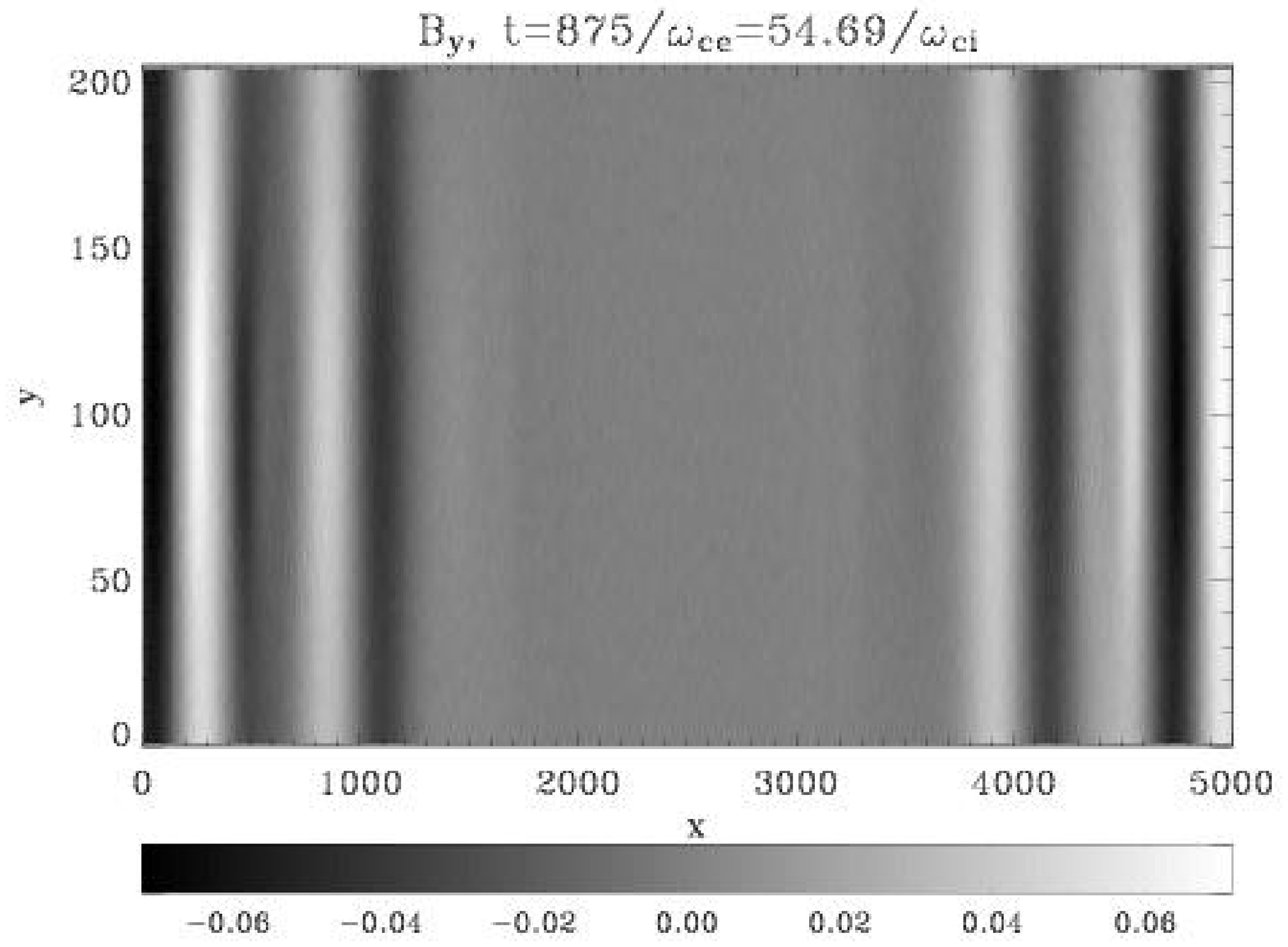,width=6cm}
 \epsfig{file=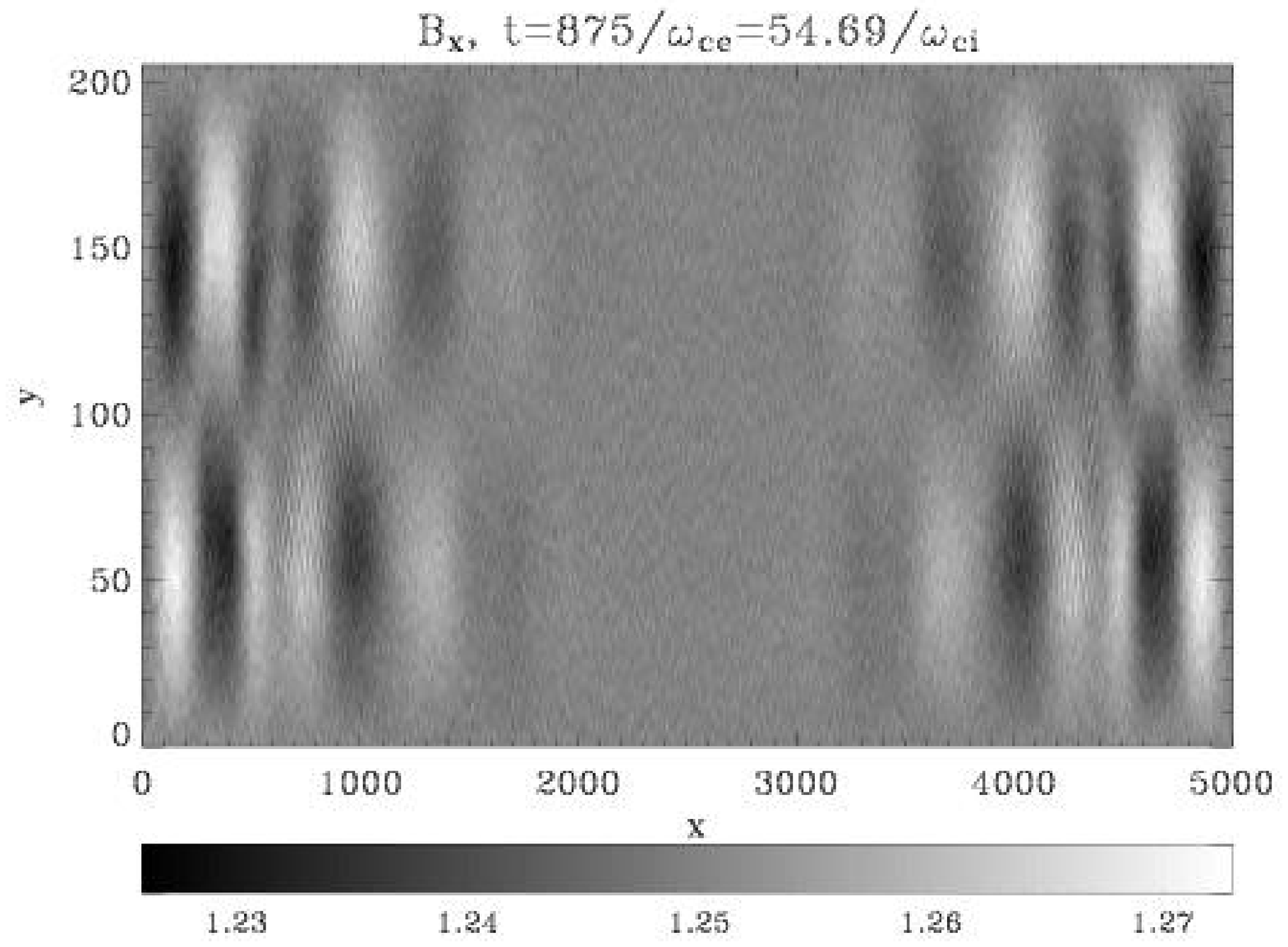,width=6cm}
 \epsfig{file=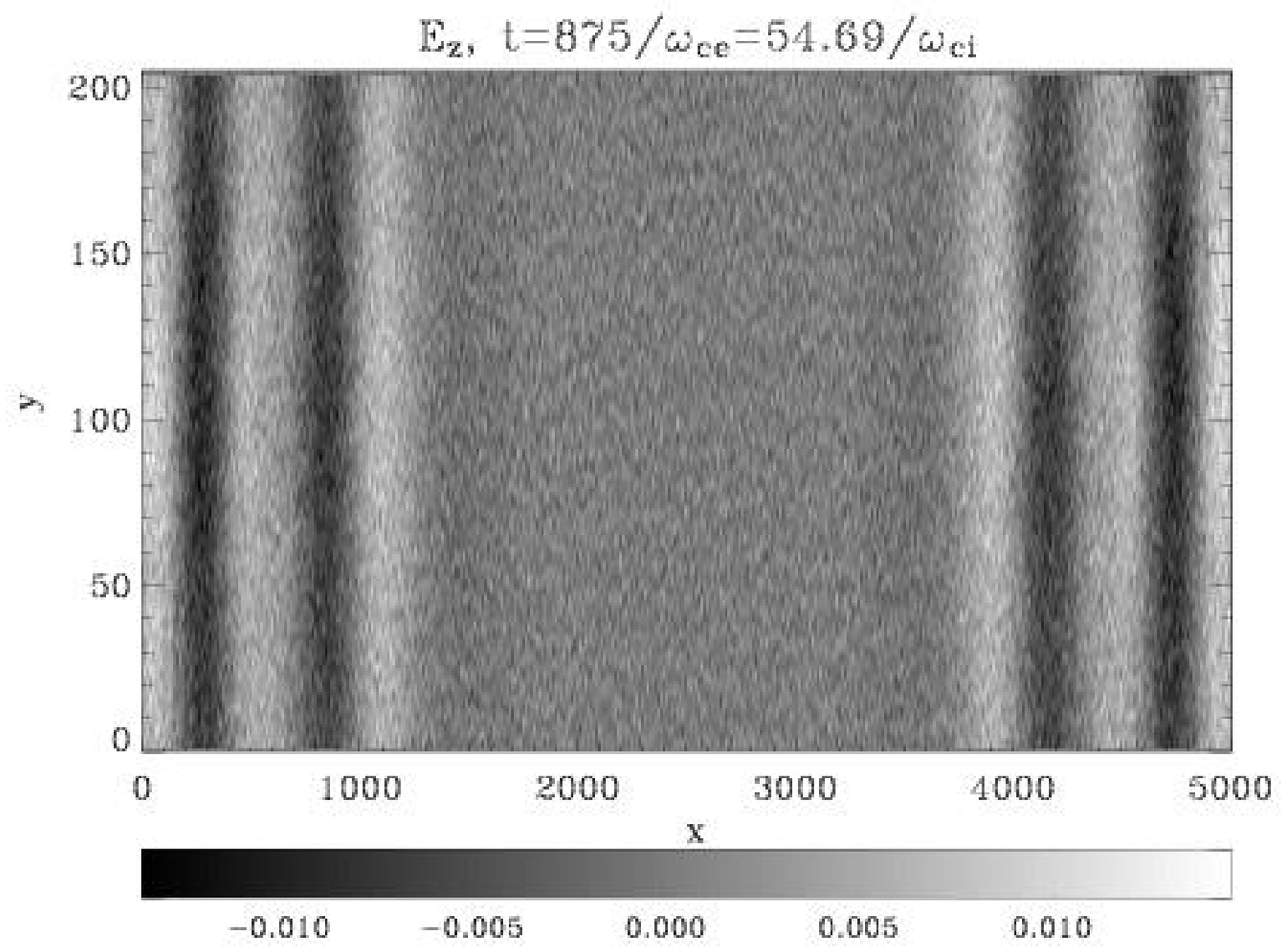,width=6cm}
 \epsfig{file=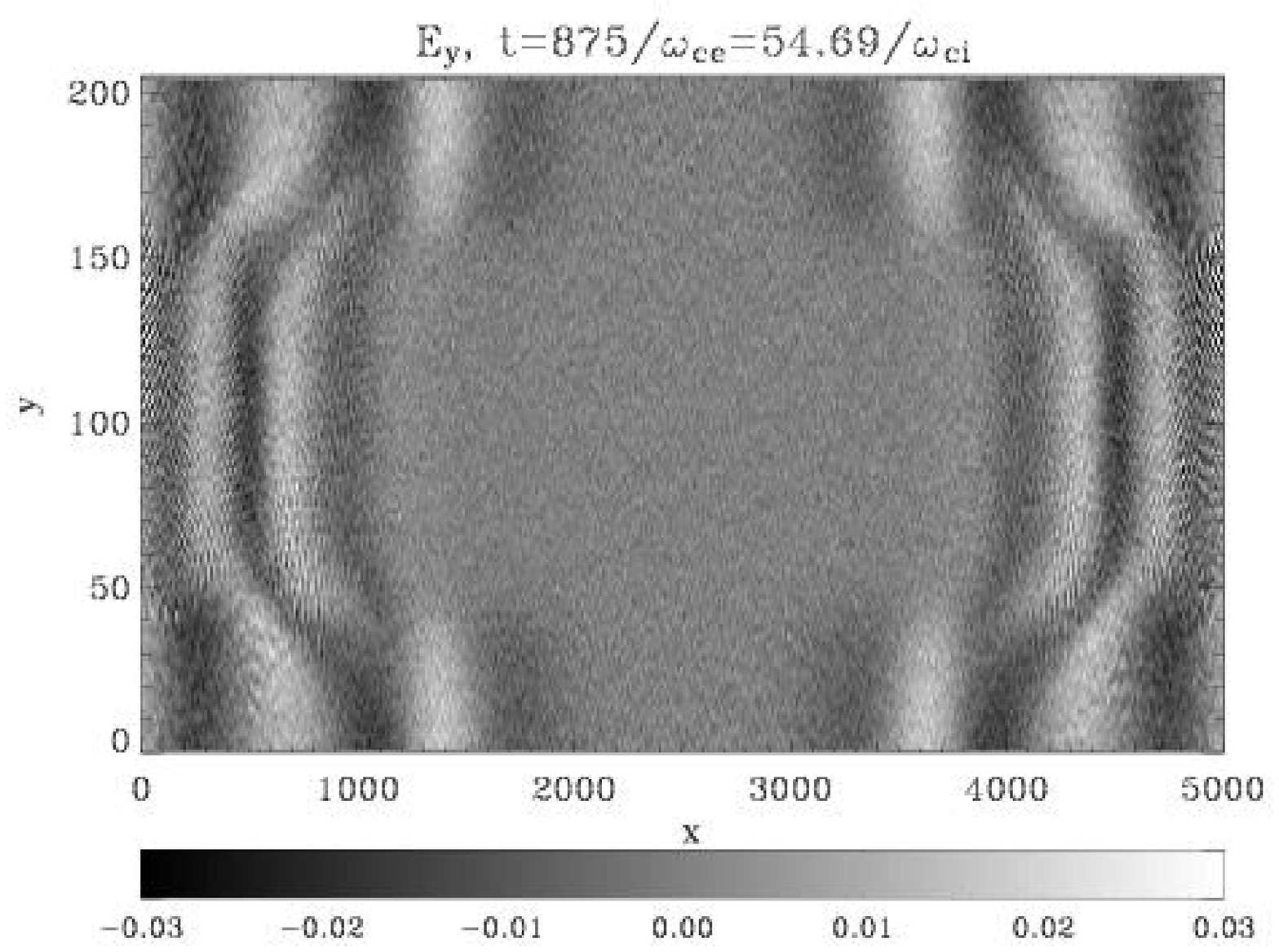,width=6cm}
 \epsfig{file=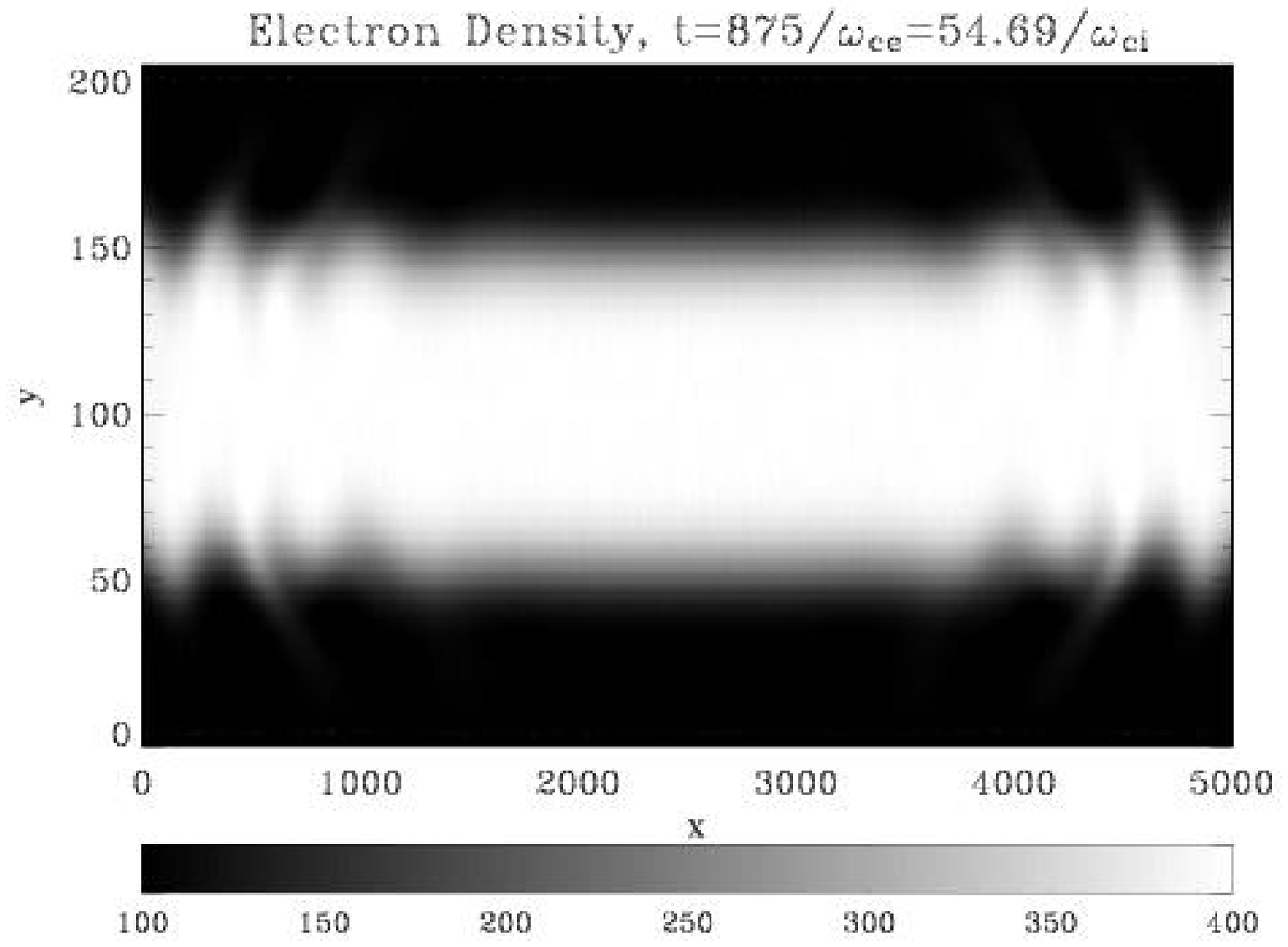,width=6cm}
\caption{Contour (intensity) plots of electromagnetic field components and electron density 
at time $t=54.69 / \omega_{ci}$ (developed stage of phase mixing). 
The phase mixed Alfv\'en wave components are $B_z$ and $E_y$. 
The excitation source is at the left
boundary. Because of periodic boundary conditions, the 
left-propagating AW re-appears from the 
right side of the simulation box. Note how the (initially plain) AW is stretched because of 
differences in local Alfv\'en speed across the 
$y$-coordinate. Significant ($\approx 10$\%) 
density fluctuations can be seen.}
\end{figure*}

In Fig.~2 we try to address the question of 
where the AW energy went? (as we saw strong
decay of AWs in the regions of strong density gradients).
Thus in Fig.~2 we plot $E_x$, the longitudinal
electrostatic field, and electron phase space ($V_x/c$ vs. $x$ and $V_x/c$ vs. $y$) for different 
times.
In the 
regions around $y \approx 50$ and $150$, for later times, a significant electrostatic field
is generated. This is the consequence of stretching of the 
AW front in those regions
because of the difference in local Alfv\'en speed.
In the middle column of this figure we see that exactly in those regions
where $E_x$ is generated, 
many electrons are accelerated along $x$-axis.
We also gather from the right column that for 
later times ($t=54.69 / \omega_{ci}$), the
number of high velocity electrons is increased around the 
strongest density gradient regions
(around $y \approx 50$ and $150$). Thus, the generated $E_x$ field is somewhat oblique
(not exactly parallel to the external magnetic field).
Hence, we conclude that the 
energy of the phase-mixed AW goes into acceleration of electrons.
Line plots of $E_x$ show that this electrostatic field is strongly damped,
i.e. the energy is channelled to electrons via Landau damping.

\begin{figure*}
\centering
\includegraphics[width=12cm]{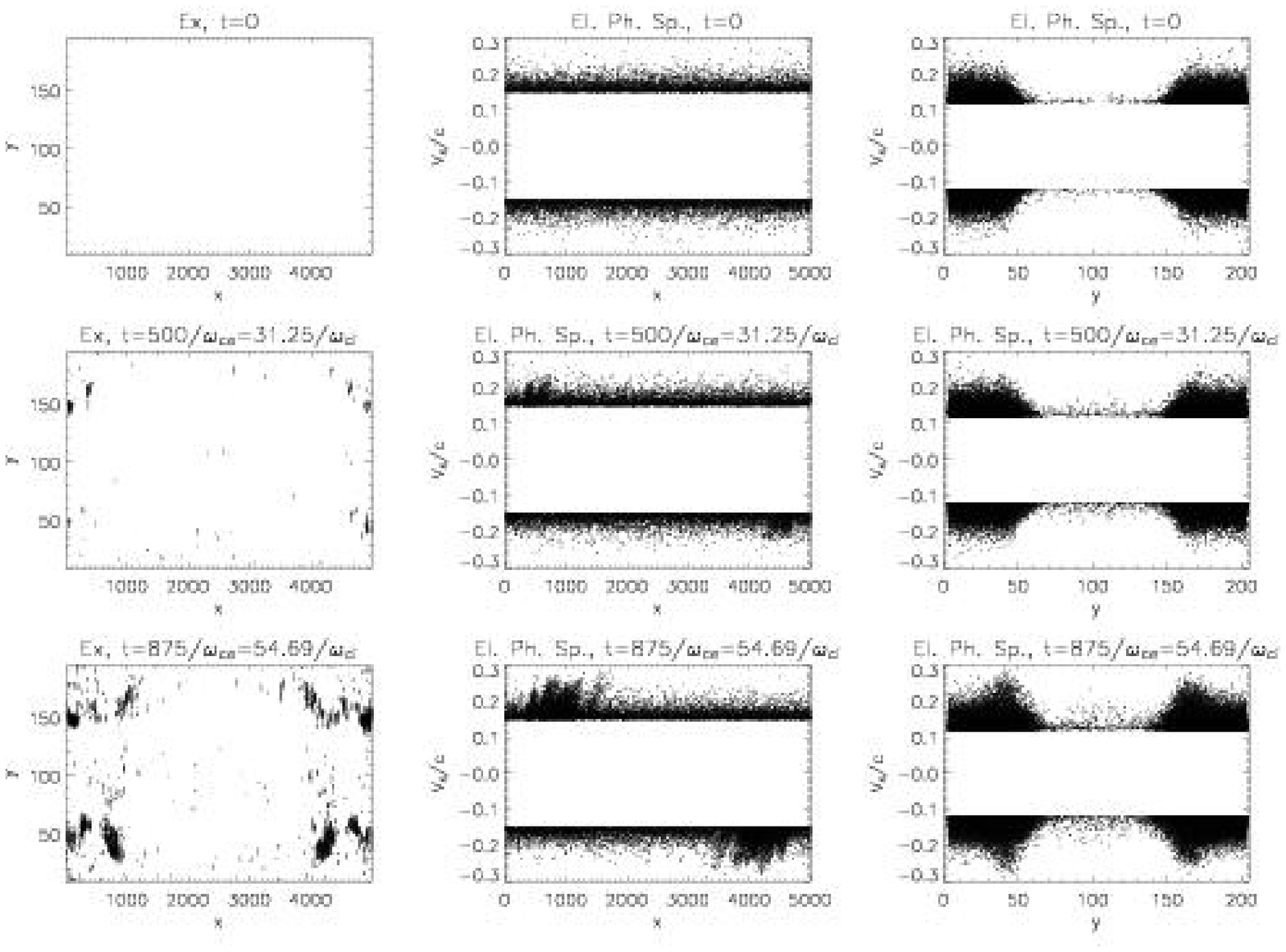}
\caption{Left column: contour plots of the generated electrostatic field $E_x$
nearly parallel to the
external magnetic field  at instances: 
$t=(0, 31.25, 54.69) / \omega_{ci}$. Central column: $V_x/c$ versus $x$ of electron phase
space at the same times. To reduce figure size, only electrons with $V_x > 0.15c$
were plotted. Right column: $V_x/c$ versus $y$ of electron phase
space at the same times. Only electrons with $V_x > 0.12c$
were plotted (note the dip in the middle due to the 
density inhomogeneity across $y$-coordinate).}
\end{figure*}

In Fig.~3 we investigate ion phase space
($V_z/c$ vs. $x$ and $V_z/c$ vs. $y$) for the different 
times. The reason for choice of $V_z$ will become clear below.
We gather from this plot that in $V_z/c$ vs. $x$ phase space, clear
propagating oscillations are present (left column). These oscillations 
are of the incompressible, Alfv\'enic "kink" type, 
i.e. for those $x$s where there is an increase of
greater positive velocity ions, there is also a 
corresponding decrease of lower negative velocity ions.
In the $V_z/c$ vs. $y$ plot we also see no clear acceleration of
ions.

\begin{figure*}
\centering
\includegraphics[width=12cm]{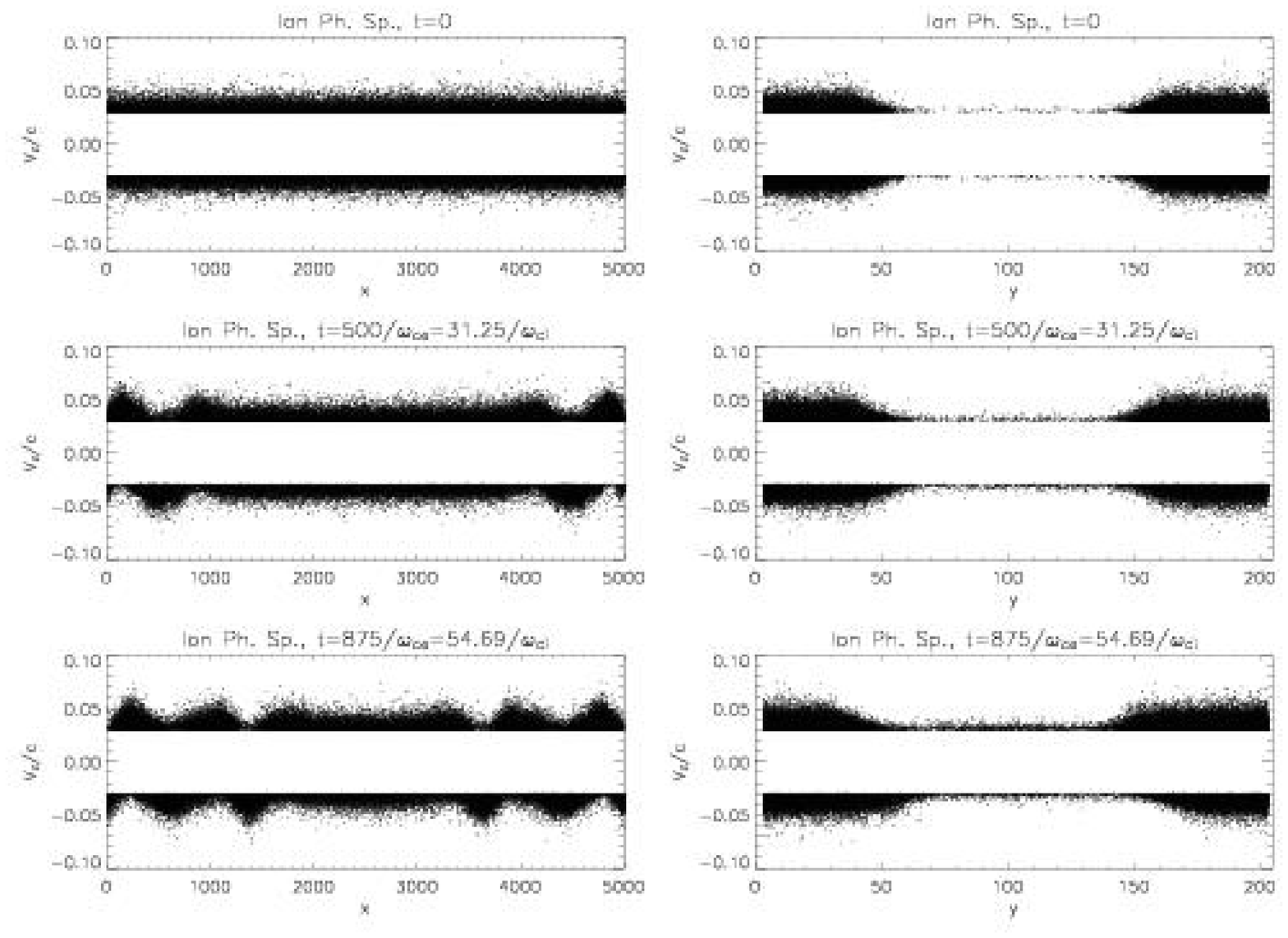}
\caption{Left column: $V_z/c$ versus $x$ of ion phase
space at instances: 
$t=(0, 31.25, 54.69) / \omega_{ci}$. 
Right column: $V_z/c$ versus $y$ of ion phase
space at the same times. Only ions with $V_z > 0.03c$
are plotted (note the dip in the middle due to the 
density inhomogeneity across the $y$-coordinate).}
\end{figure*}

We next look at the distribution functions of electrons and ions
before and after the phase mixing took place.
In Fig.~4 we plot distribution functions of electrons and ions at $t=0$ and $t=54.69 / \omega_{ci}$.
Note that  at $t=0$  the distribution functions do not look as 
purely Maxwellian because
of the fact that the 
temperature varies across the $y$-coordinate (to keep total pressure
constant) and the graphs are produced for the entire simulation domain.
Also, note that for electrons in $f(V_x)$ there is 
a substantial difference at $t=54.69 / \omega_{ci}$
to its original form because of the aforementioned 
electron acceleration.
We see that the number of electrons having velocities $V_x=\pm (0.1-0.3)c$ is increased.
Note that the acceleration of electrons takes place mostly along
the external magnetic field (along the $x$-coordinate). Thus, very little electron acceleration 
occurs for $V_y$ or $V_z$ (solid and dotted curves practically overlap each other).
For the ions the situation is different: we see broadening of the ion velocity distribution functions
in $V_z$  and $V_y$ (that is why we have chosen to present the 
$V_z$ component of ion phase space in Fig.~3).
\begin{figure*}
\centering
\includegraphics[width=12cm]{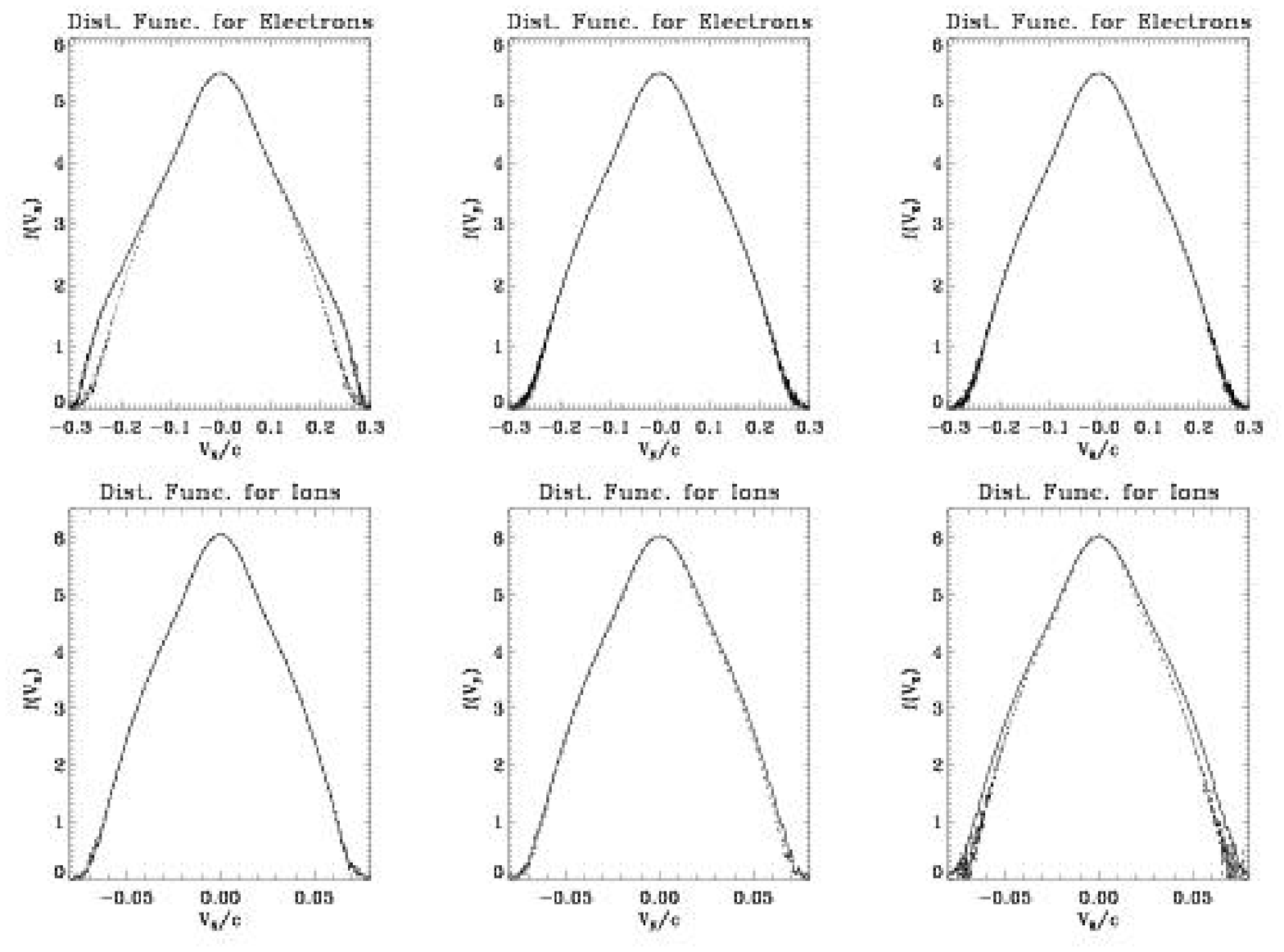}
\caption{All three components of the distribution functions of electrons 
(top row) and ions (bottom row) 
at $t=0$ (dotted curves) and $t=54.69 / \omega_{ci}$
(solid curves), i.e. for the developed stage of phase mixing.}
\end{figure*}

The reason for this broadening of the ion distribution function becomes 
clear in Fig.~5 where we plot
kinetic energy $x,y,z$ components ($\propto V_{x,y,z}^2$) and total kinetic energies for
electrons (top row) and ions (bottom row). For ions we gather that $y$ and $z$ components of
the kinetic energy (bottom left figure) oscillate in anti-phase and their oscillatory part perfectly cancels 
out in the total energy (bottom right figure). Thus, the broadening of the $y$ and $z$ components of the
ion velocity distribution functions is due to the presence of AWs (usual wave broadening, which is actually observed
e.g. in the solar corona and solar wind, \cite{bd71,sbnm95,bpb00}), and hence there is no ion acceleration present.
Note that $y$ and $z$ components and hence total kinetic energy of ions is monotonously increasing due to continuous
AW driving. Note that no significant motion of ions along the field is present.
For electrons, on the other hand, we see a 
significant increase of the 
$x$ component (along the magnetic field) of kinetic energy
which is due to the new electron acceleration mechanism discovered by us (cf. Fig.10).
Note that for ions the $y$ component reaches lower values (than the 
$z$ component) because of lower AW velocity in the middle 
part of the simulation domain.

\begin{figure}[]
\resizebox{\hsize}{!}{\includegraphics{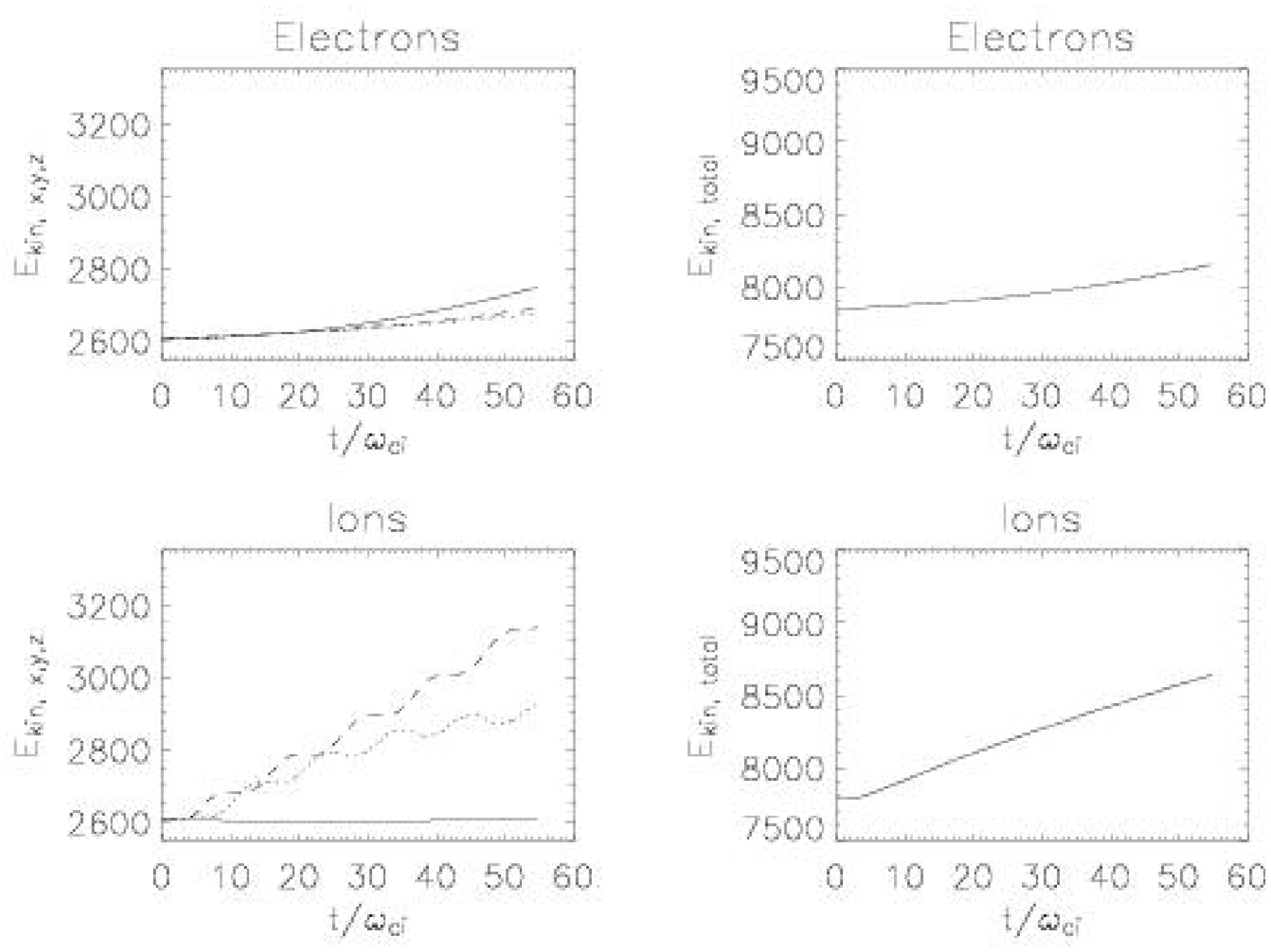}} 
\caption{Top row: kinetic energies (calculated  by $x$ (solid), $y$ (dotted), $z$ (dashed) velocity (squared) components) of 
electrons (left), and total kinetic energy for electrons (right)
as a function of time. Bottom row: As above but for ions. The units on $y$-axis are arbitrary.}
\end{figure}

The next step is to check whether the increase in electron velocities really comes from the
resonant wave particle interactions. For this purpose in Fig.~6, left 
panel,  we plot two snapshots of the
Alfv\'en wave $B_z(x,y=148)$ component at instances $t=54.69 / \omega_{ci}$ (solid line)
and $t=46.87 / \omega_{ci}$ (dotted line).
The distance between the two upper leftmost peaks (which is the distance 
travelled by the
wave in the time span between the snapshots) 
is about $\delta L=150\Delta=15(c/\omega_{pe})$.
The time difference between the snapshots is $\delta t=7.82 / \omega_{ci}$.
Thus, the 
measured AW speed at the point of the strongest density gradient ($y=148$)
is $V_A^M=\delta L /\delta t=0.12c$. We can also work out 
the Alfv\'en speed from theory.
In the homogeneous low density region the Alfv\'en speed was set to be
$V_A(\infty)=0.25 c$. From Eq.(1) it follows that for $y=148$ the 
density is increased by a factor of
$2.37$ which means that the Alfv\'en wave speed at this position is
$V_A(148)=0.25/\sqrt{2.37}c=0.16c$.
The measured ($0.12c$) and calculated ($0.16c$) Alfv\'en speeds in the inhomogeneous regions 
do not coincide. This is probably because  the 
AW front is decelerated (due to momentum conservation) 
as it passes on energy and momentum to the
electrons in the inhomogeneous regions (where electron acceleration takes place). 
However, this possibly is not the case if wave-particle interactions
play the same role as dissipation in  MHD \citep{sg69}:
Then wave-particle interactions would result only in the decrease of the AW
amplitude (dissipation) and not in its deceleration.
If we compare these values to Fig.~4 (top left panel for $f(V_x)$), we deduce
 that these are the
velocities $>0.12c$ above which electron numbers with higher velocities
are greatly increased. This deviation peaks at about $0.25c$ which,
in fact, corresponds to the Alfv\'en speed in the lower density regions.
This can be explained by the fact the electron acceleration takes
place in  wide regions (cf. Fig.~2) along and around $y \approx 148$ (and $y \approx 51$) -- hence
the spread in the accelerated velocities.
In Fig.~6 we also plot a visual fit curve (dashed line) to 
quantify the amplitude decay law for the AW (at $t=54.69 / \omega_{ci}$)
in the strongest density inhomogeneity region.
The fitted (dashed) cure is represented by $0.056 \exp \left[ -
\left({x}/{1250}\right)^3\right]$.
There is a surprising similarity of this fit with the 
MHD approximation results.
\citet{hp83} found that for large times (developed phase mixing),
in the case of a harmonic driver, the amplitude decay law
is given by $\propto \exp \left[ -
\left(\frac{\eta \omega^2 V_A^{\prime 2}}{6 V_A^{5}}\right)x^3\right]$ which 
is much faster 
than the usual resistivity dissipation
$\propto \exp(-\eta x)$. Here $V_A^{\prime}$ is the derivative
of the Alfv\'en speed with respect to the $y$-coordinate.
The most interesting fact is that even in the kinetic approximation
the same $\propto \exp (-A x^3)$ law holds as in MHD.

\begin{figure*}
\centering
\includegraphics[width=12cm]{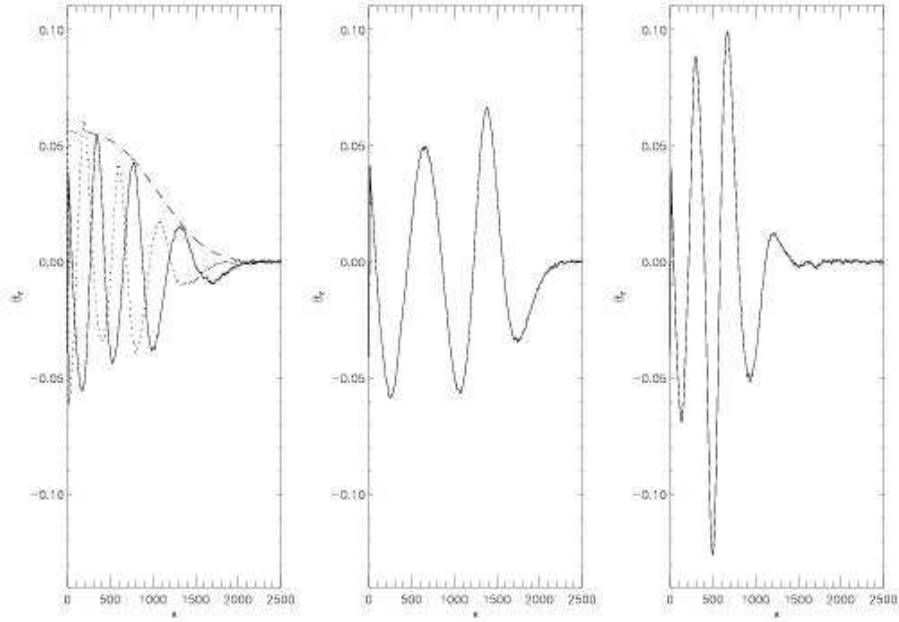}
\caption{Left: two snapshots of the
Alfv\'en wave $B_z(x,y=148)$ component at instances $t=54.69 / \omega_{ci}$ (solid line)
and $t=46.87 / \omega_{ci}$ (dotted line). The dashed line represents fit
$0.056 \exp \left[ -\left({x}/{1250}\right)^3\right]$. Center: $B_z(x,y=10)$ 
(low density homogeneous region), $B_z(x,y=100)$ (high density homogeneous region).
Note the
differences in amplitudes and propagation speeds, which are consistent with the equilibrium
density and thus Alfv\'en speed dependence on $y$-coordinate.}
\end{figure*}

In MHD, finite resistivity and
Alfv\'en speed non-uniformity are responsible for the
enhanced dissipation via phase mixing.
In our PIC simulations (kinetic phase mixing), however, we do not have dissipation
and collisions (dissipation). Thus, in our case,
wave-particle interactions play the same role as 
resistivity $\eta$ in the MHD phase mixing \citep{sg69}.
No significant AW dissipation
was found away from the density inhomogeneity regions (Fig.~6 middle and right panels, note also the
differences in amplitudes and propagation speeds, which are consistent with the imposed density and hence Alfv\'en speed variation
across the $y$-coordinate).
This has the same explanation as in the case of MHD --
it is in the regions of density of inhomogeneities ($V_A^{\prime}\not=0$) 
that the dissipation is greatly enhanced, while in the regions
where $V_A^{\prime}=0$ there is no substantial dissipation (apart from the 
classical $\propto \exp(-\eta x)$
one).
In the MHD approximation, the aforementioned amplitude decay law is derived
from the diffusion equation, to which MHD equations reduce for large times (developed
phase mixing \citep{tnr03}). It seems that the kinetic description 
leads to the same type of diffusion equation.
It is unclear at this stage, however, what physical quantity 
would play the role of resistivity $\eta$ (from the MHD approximation) in the
kinetic regime.

\subsection{Homogeneous plasma case}

In order to clarify the broadening of the ion velocity distribution function
 and also for a consistency check
we performed an additional simulation in the case of homogeneous plasma.
Now the density was fixed at 100 ions/electrons per cell in the entire simulation
domain and hence plasma temperature and thermal velocities were fixed too.
In such a set up no phase mixing should take place as the AW speed is uniform.

\begin{figure*}
\centering
 \epsfig{file=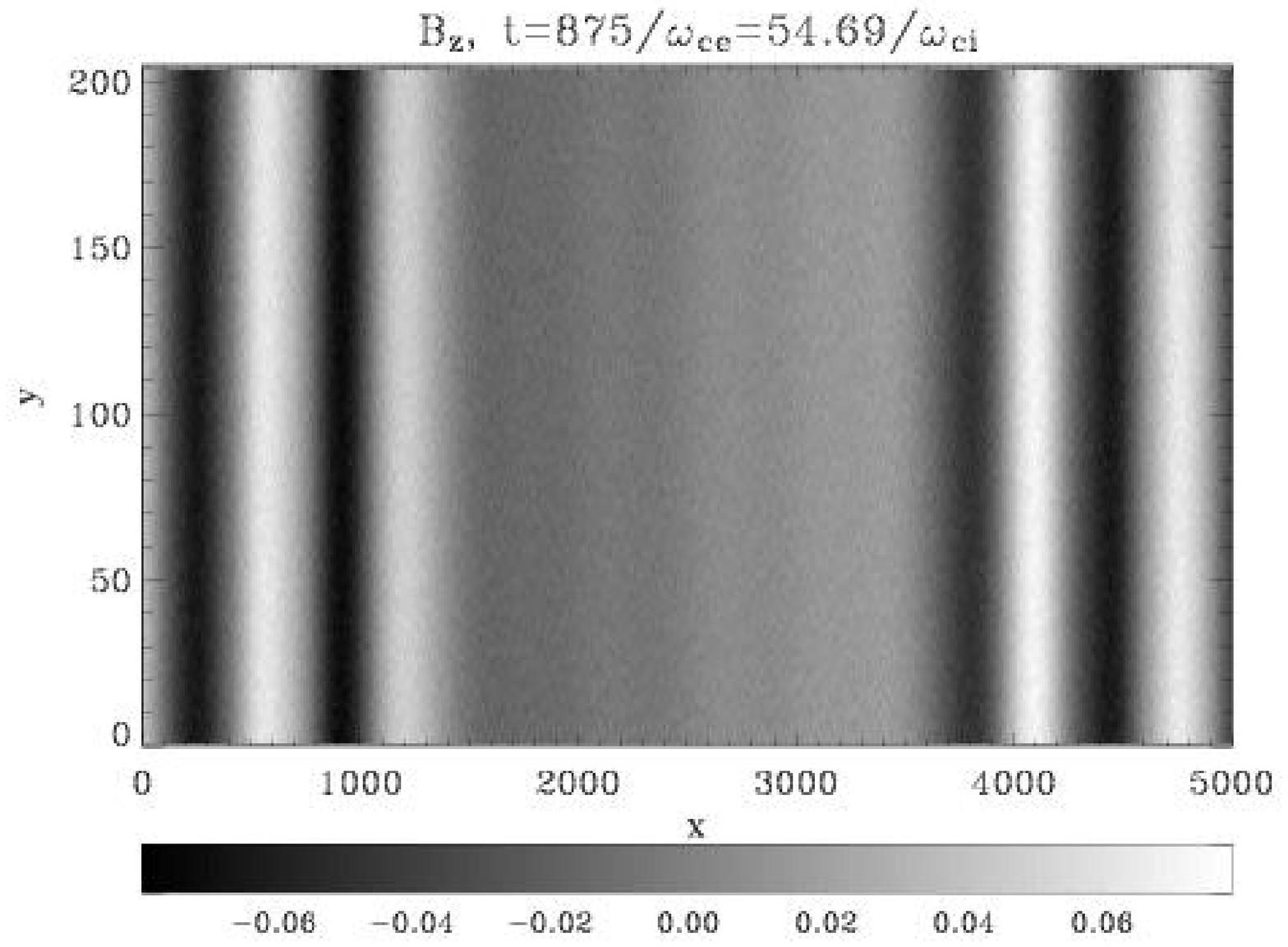,width=6cm}
 \epsfig{file=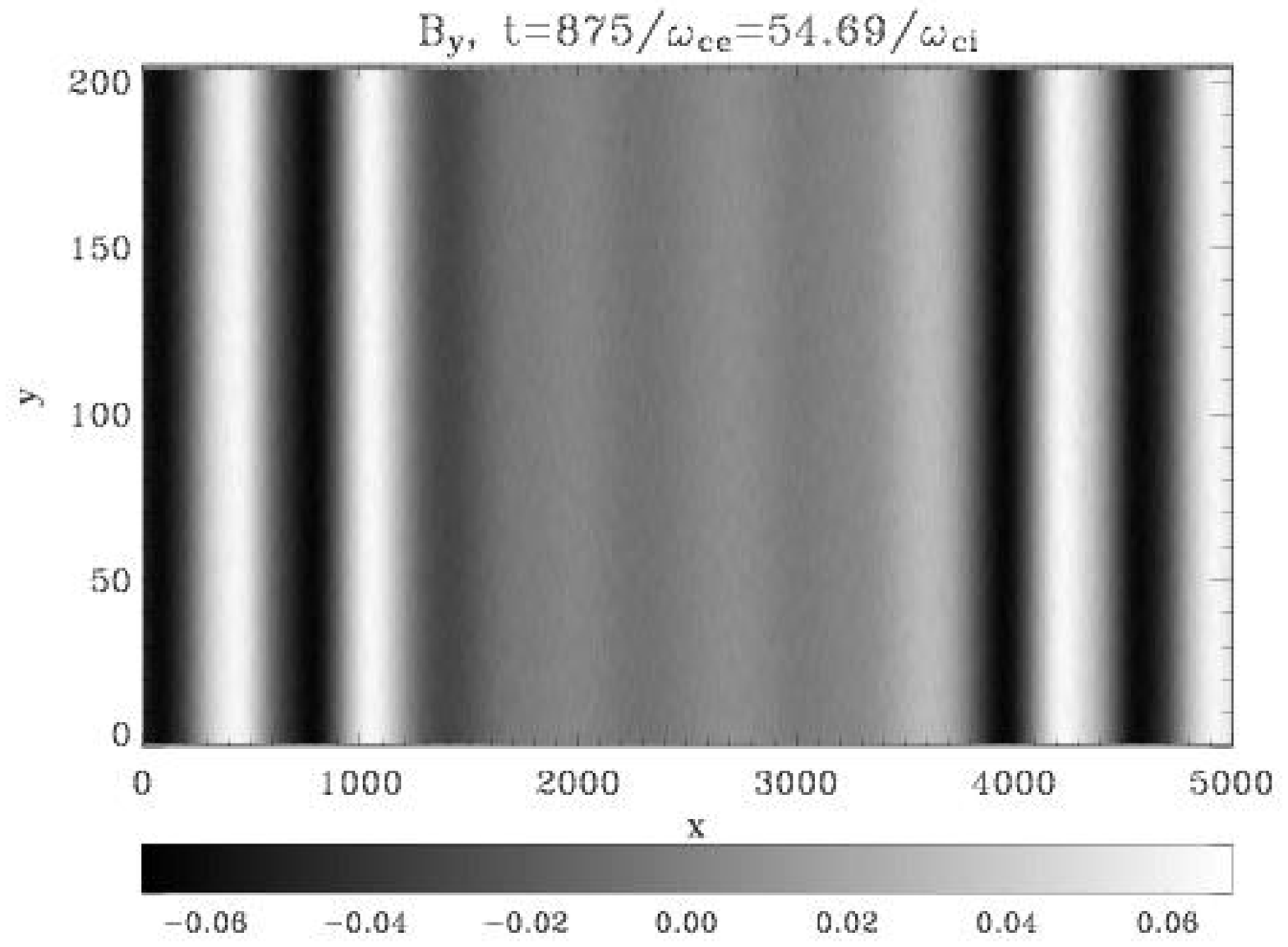,width=6cm}
 \epsfig{file=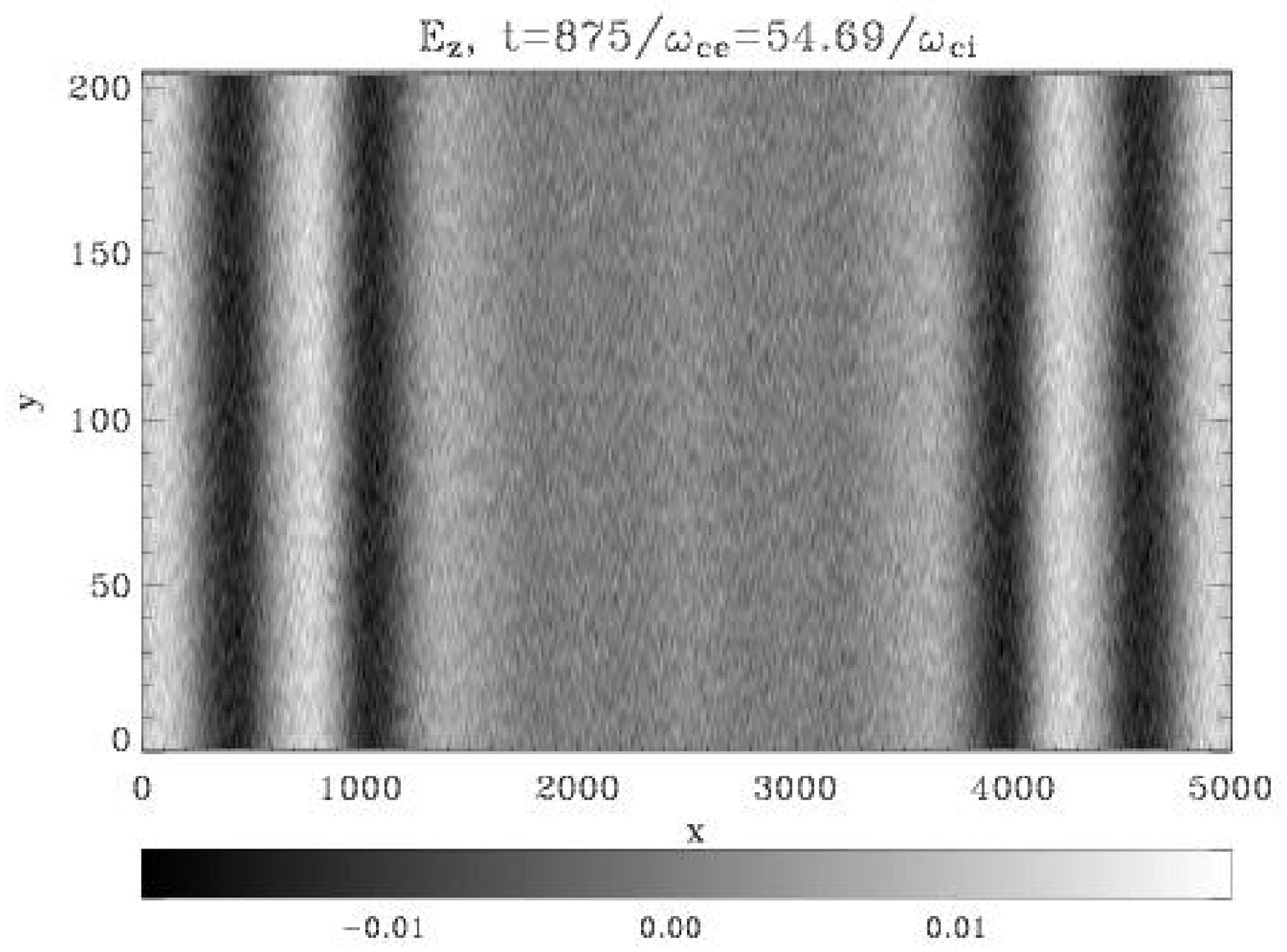,width=6cm}
 \epsfig{file=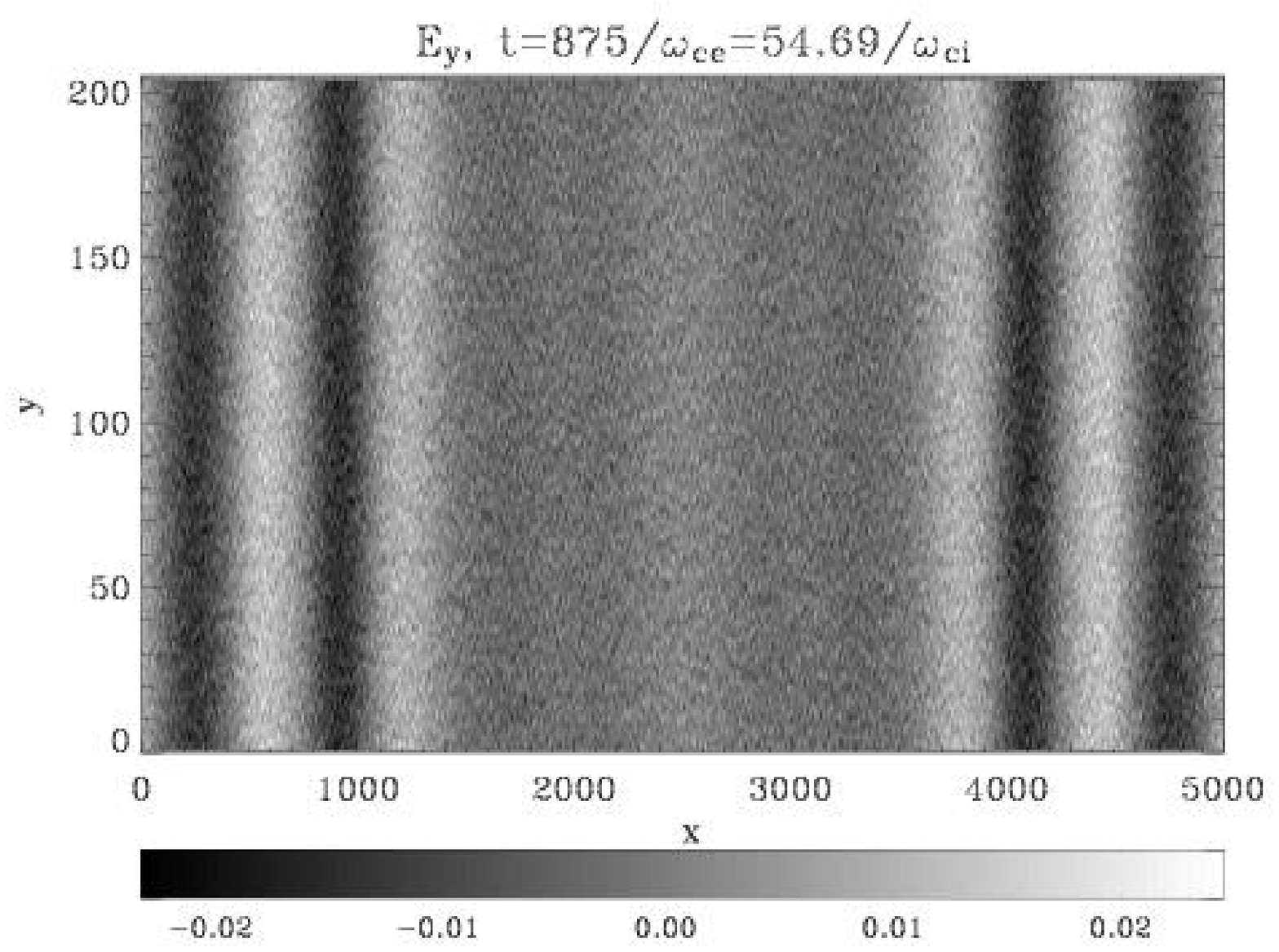,width=6cm}
 \caption{As in Fig.~1 but for the case of homogeneous plasma density (no phase mixing).
 Note only non-zero (above noise level) components are plotted. There are no $B_x,E_x$ or density
 fluctuations present in this case.}
\end{figure*}

In Fig.~7 we plot the only non-zero (above noise level) components at $t=54.69 / \omega_{ci}$, which
are left circularly polarised AW fields: $B_z,B_y,E_z,E_y$. 
Note there are no $B_x,E_x$ or density fluctuations present in this case (cf. Fig.~1) as
it is the plasma inhomogeneity that facilitates the coupling between  AW and the compressive
modes.
\begin{figure*}
\centering
\includegraphics[width=12cm]{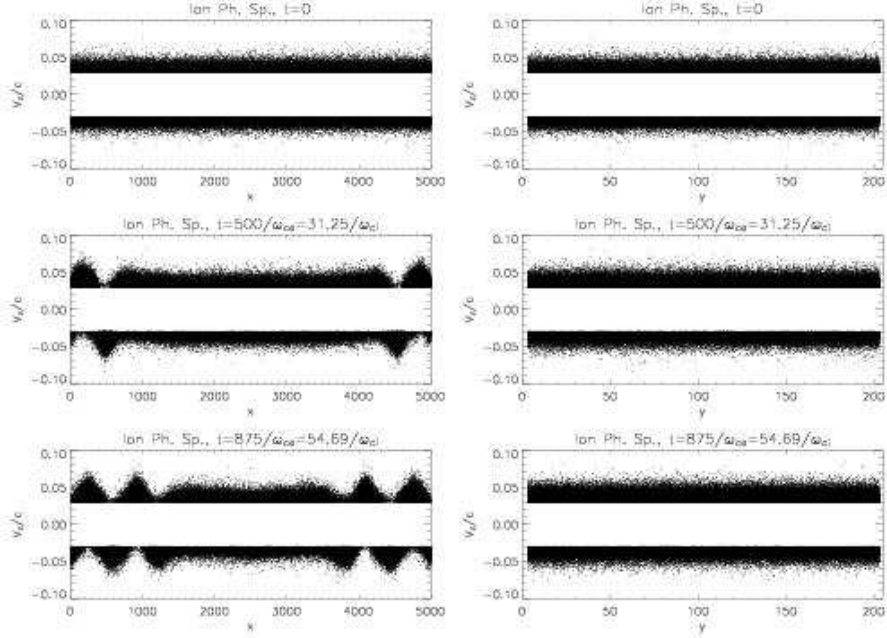}
\caption{As in Fig.~3 but for the case of homogeneous plasma density (no phase mixing).}
\end{figure*}

In Fig.~8 we plot ion phase space ($V_z/c$ vs. $x$ and $V_z/c$ vs. $y$) in the
homogeneous plasma case for different times (cf. Fig.~3). We gather from the graph that
propagating, incompressible, Alfv\'enic "kink" type oscillations are still present (left column),
while no significant ion
acceleration takes place (right column). This is better understood from Fig.~9 where we plot electron and ion 
distribution functions for $t=0$ and $t=54.69 / \omega_{ci}$ (as in Fig.~4) for the 
homogeneous plasma case. 
There are three noteworthy points: (i) no electron acceleration takes place because of
the absence of phase mixing; (ii) there is (as in the inhomogeneous case) broadening of the ion
velocity distribution functions (in $V_y$ and $V_z$) due to the present AW (wave broadening);
(iii) The distribution now looks Maxwellian (cf. Fig.~4) because 
the distribution function
holds for the entire homogeneous region.

\begin{figure*}
\centering
\includegraphics[width=12cm]{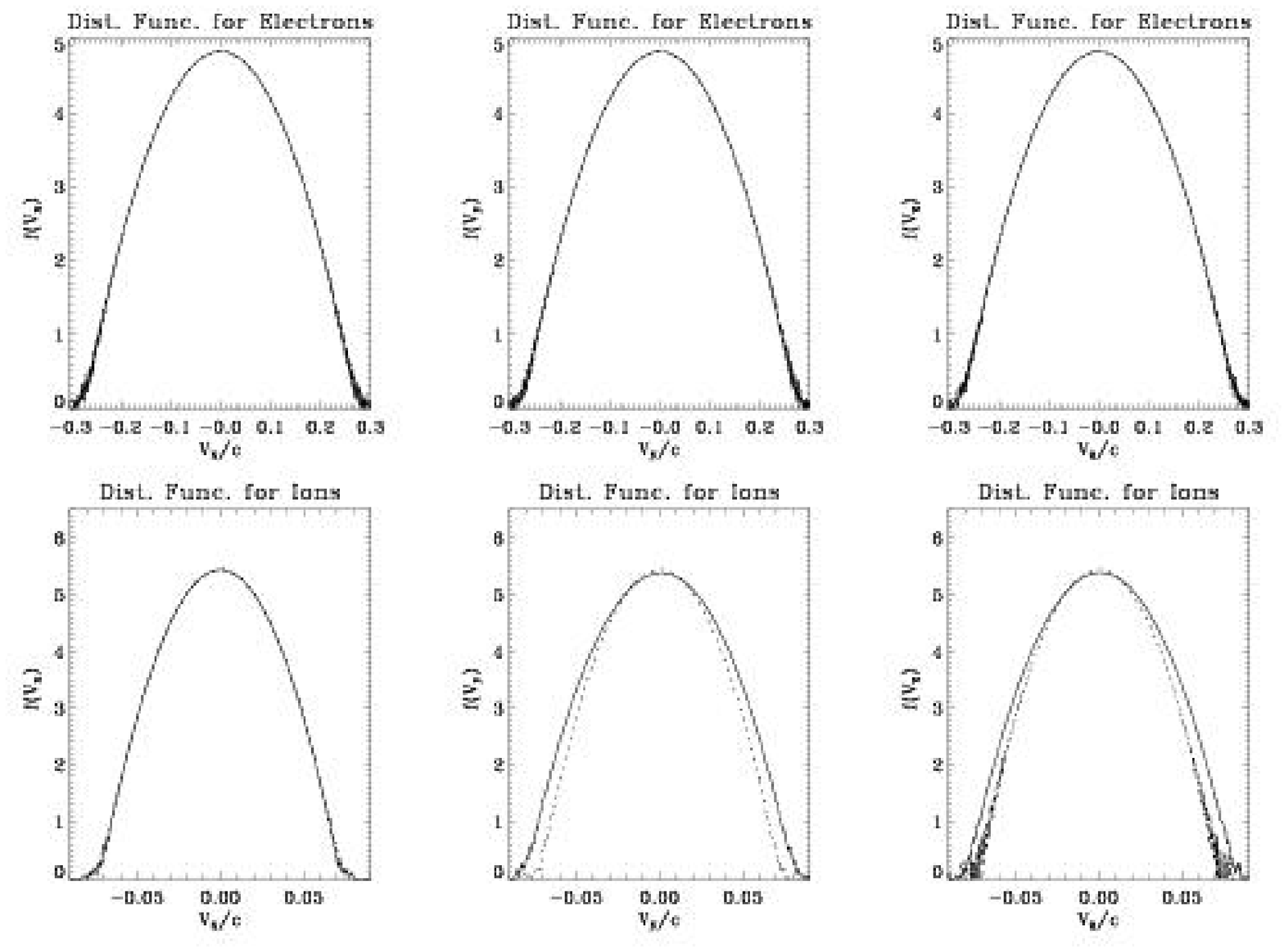}
\caption{As in Fig.~4 but for the case of homogeneous plasma density (no phase mixing).}
\end{figure*}

In Fig.~10  we plot the
kinetic energy $x,y,z$ components ($\propto V_{x,y,z}^2$) and 
total kinetic energy for
electrons (top row) and ions (bottom row). For ions we see (as in the inhomogeneous case) 
that $y$ and $z$ components of
the kinetic energy (bottom left figure) oscillate in anti-phase and their oscillatory part perfectly cancels 
out in the total energy (bottom right figure). Thus, the broadening of the $y$ and $z$ components of the
ion velocity distribution functions is due to the presence of AWs, and, in turn, there is no ion acceleration.
Again the $y$ and $z$ components and hence the 
total kinetic energy of the ions is monotonously increasing due to continuous
AW driving. No significant motion of ions along the field is present.
For electrons we do not observe any acceleration due to the absence of phase mixing (cf. Fig.5).
Note that the $y$ component now attains the same values as the 
$z$ component (bottom left figure)
because of the same AW velocity in the entire simulation domain.

\begin{figure}[]
\resizebox{\hsize}{!}{\includegraphics{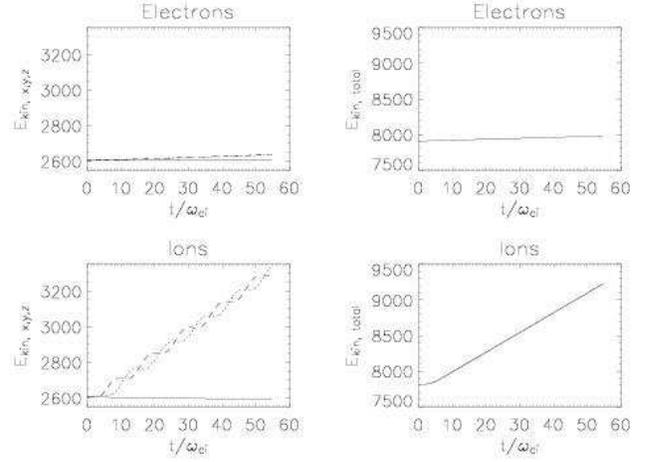}} 
\caption{As in Fig.~5 but for the case of homogeneous plasma density (no phase mixing).} 
\end{figure}

\section{Discussion}

In our preliminary work \citep{tss05} we outlined the main results of our
newly-discovered mechanism of electron acceleration. Here we 
presented a more detailed analysis
of the phenomenon. We have established the following:

\begin{itemize}
\item Progressive distortion of the Alfv\'en wave front, due to the differences in 
local Alfv\'en speed, generates oblique (nearly parallel to the magnetic
field) electrostatic fields, which accelerate electrons.

\item The amplitude decay law in the inhomogeneous regions, 
in the kinetic regime, is shown to be the same as in the MHD approximation 
described by \citet{hp83}.

\item The density perturbations ($\approx 10$\% of background)
are generated due to both the weak non-linearity and plasma inhomogeneity. 
These are propagating density oscillations with variations both 
in overall magnitude 
and across the $y$-coordinate. They are mainly confined to the strongest density gradients regions
(around $y \approx 50$ and $150$) i.e. edges of the density structure (e.g. boundary of a solar coronal loop). 
Longitudinal to the external magnetic field, $B_x$, perturbations are also generated in the
same manner, but with smaller ($\approx 3$\%) amplitudes.

\item Both in the homogeneous and inhomogeneous cases the presence 
of AWs causes broadening of the perpendicular 
(to the external magnetic field) ion velocity 
distribution functions, while no ion acceleration is observed.

\end{itemize}

In the MHD approximation 
\citet{hbw02} and \citet{tnr03} showed that 
in the case of localised Alfv\'en pulses,
Heyvaerts and Priest's amplitude decay
formula $\propto \exp (-A x^3)$ (which is true for
harmonic AWs) is replaced by the power law $B_z \propto x^{-3/2}$. 
A natural next step forward
would be to check whether 
in the case of localised Alfv\'en pulses the same power law holds
in the kinetic regime.

After this study was complete
we became aware of a study by \citet{vh04}, who used a hybrid
code (electrons treated as a neutralising fluid, with ion
kinetics retained) as opposed to our (fully kinetic) PIC code,
to simulate resonant absorption. They found that 
a planar (body) Alfv\'en
wave propagating at less than $90^{\circ}$ to a background gradient  
has field lines which lose wave energy to another set of field lines by
cross-field transport. Further, \citet{v04} found that 
when perpendicular scales of the
order of 10 proton inertial lengths ($10 c/\omega_{pi}$) 
develop from wave refraction
in the vicinity of the resonant field lines, a non-propagating 
density fluctuation begins
to grow to large amplitudes. This saturates by exciting highly 
oblique, compressive and
low-frequency waves which dissipate and heat protons. 
These processes lead to a faster development of small
scales across the magnetic field, i.e. the phase mixing mechanism, 
studied here.

\begin{acknowledgements}
The authors gratefully acknowledge support from
CAMPUS (Campaign to Promote University of Salford) which funded
J.-I.S.'s one month fellowship to the Salford University 
that made this project possible.
DT acknowledges use of E. Copson Math cluster 
funded by PPARC and University of St. Andrews.
DT kindly acknowledges support from Nuffield Foundation 
through an award to newly appointed lecturers in Science,
Engineering and Mathematics (NUF-NAL 04).
The authors would like to 
thank the referee, Dr. Bernie J. Vasquez, for pointing out some minor
inconsistencies, which have been now corrected.
\end{acknowledgements}

\bibliography{ms2436}% Produces the bibliography via BibTeX.
\end{document}